\documentclass[12pt]{article}
\usepackage{cite,epsfig}

\catcode`@=11
\def\citer{\@ifnextchar [{\@tempswatrue\@citexr}{\@tempswafalse\@citexr[]}}
 
\def\@citexr[#1]#2{\if@filesw\immediate\write\@auxout{\string\citation{#2}}\fi
  \def\@citea{}\@cite{\@for\@citeb:=#2\do
    {\@citea\def\@citea{--\penalty\@m}\@ifundefined
       {b@\@citeb}{{\bf ?}\@warning
       {Citation `\@citeb' on page \thepage \space undefined}}%
\hbox{\csname b@\@citeb\endcsname}}}{#1}}
\catcode`@=12

\def\refeq#1{\mbox{eq.~(\ref{#1})}}
\def\refeqs#1{\mbox{eqs.~(\ref{#1})}}
\def\reffi#1{\mbox{Fig.~\ref{#1}}}

\def\refse#1{\mbox{Sect.~\ref{#1}}}

\def\citere#1{\mbox{Ref.~\cite{#1}}}
\def\citeres#1{\mbox{Refs.~\cite{#1}}}

\newcommand{\mste}{m_{\tilde{t}_1}}
\newcommand{\mstz}{m_{\tilde{t}_2}}

\newcommand{\MstL}{M_{\tilde{t}_L}}
\newcommand{\MstR}{M_{\tilde{t}_R}}

\newcommand{\Mtlr}{M_{t}^{LR}}
\newcommand{\Mtlrz}{\KL M_{t}^{LR}\KR^2}

\newcommand{\Mtlrv}{\KL M_{t}^{LR}\KR^4}

\newcommand{\Mtlrse}{\KL M_{t}^{LR}\KR^6}

\newcommand{\Mtlra}{\KL M_{t}^{LR}\KR^8}
\newcommand{\Mblr}{M_{b}^{LR}}
\newcommand{\Mblrz}{\KL M_{b}^{LR}\KR^2}

\newcommand{\ms}{M_S}
\newcommand{\msusy}{M_{SUSY}}
\newcommand{\msq}{m_{\tilde{q}}}

\newcommand{\Pe}{\phi_1}
\newcommand{\Pz}{\phi_2}

\newcommand{\PePz}{\phi_1\phi_2}
\newcommand{\mpe}{m_{\Pe}}
\newcommand{\mpz}{m_{\Pz}}
\newcommand{\mpez}{m_{\PePz}}

\newcommand{\SU}{\mathrm {SUSY}}

\newcommand{\msbar}{$\overline{\rm{MS}}$}
\newcommand{\oa}{{\cal O}(\alpha)}

\newcommand{\oaas}{{\cal O}(\alpha\alpha_s)}
\newcommand{\cp}{{\cal CP}}
\newcommand{\wz}{\sqrt{2}}
\newcommand{\edz}{\frac{1}{2}}
\newcommand{\twol}{two-loop}
\newcommand{\onel}{one-loop}

\newcommand{\fh}{{\em FeynHiggs}}
\newcommand{\fhf}{{\em FeynHiggsFast}}

\newcommand{\MW}{M_W}
\newcommand{\MZ}{M_Z}
\newcommand{\MA}{M_A}
\newcommand{\mh}{m_h}

\newcommand{\dr}{\De\rho}
\newcommand{\mt}{m_{t}}
\newcommand{\mtms}{\overline{m}_t}

\newcommand{\mb}{m_{b}}

\newcommand{\mgl}{m_{\tilde{g}}}

\newcommand{\Stop}{\tilde{t}}
\newcommand{\StopL}{\tilde{t}_L}
\newcommand{\StopR}{\tilde{t}_R}
\newcommand{\Stope}{\tilde{t}_1}
\newcommand{\Stopz}{\tilde{t}_2}

\newcommand{\Sbot}{\tilde{b}}

\newcommand{\dst}{\Delta_{\tilde{t}}}
\newcommand{\tst}{\theta_{\tilde{t}}}

\newcommand{\tsf}{\theta\kern-.20em_{\tilde{f}}}
\newcommand{\tsfp}{\theta\kern-.20em_{\tilde{f}\prime}}
\newcommand{\tsq}{\theta\kern-.15em_{\tilde{q}}}
\newcommand{\sw}{s_W}

\newcommand{\sintt}{\sin\tst}

\newcommand{\sinQZtt}{\sin^2 2\tst}

\newcommand{\costt}{\cos\tst}

\newcommand{\KL}{\left(}
\newcommand{\KR}{\right)}
\newcommand{\KKL}{\left[}
\newcommand{\KKR}{\right]}
\newcommand{\KKKL}{\left\{}
\newcommand{\KKKR}{\right\}}

\newcommand{\VL}{\left( \begin{array}{c}}
\newcommand{\VR}{\end{array} \right)}
\newcommand{\ML}{\left( \begin{array}{cc}}
\newcommand{\MLd}{\left( \begin{array}{ccc}}
\newcommand{\MLv}{\left( \begin{array}{cccc}}
\newcommand{\MR}{\end{array} \right)}

\newcommand{\tb}{\tan \beta}
\newcommand{\TQb}{\tan^2 \beta\hspace{1mm}}
\newcommand{\CTb}{\cot \beta\hspace{1mm}}

\newcommand{\Sb}{\sin \beta\hspace{1mm}}
\newcommand{\sbe}{\sin \beta}
\newcommand{\SQb}{\sin^2\beta\hspace{1mm}}

\newcommand{\SVb}{\sin^4\beta\hspace{1mm}}
\newcommand{\Cb}{\cos \beta\hspace{1mm}}
\newcommand{\CQb}{\cos^2\beta\hspace{1mm}}

\newcommand{\CVb}{\cos^4\beta\hspace{1mm}}

\newcommand{\CZb}{\cos 2\beta\hspace{1mm}}
\newcommand{\CQZb}{\cos^2 2\beta\hspace{1mm}}

\newcommand{\CVZb}{\cos^4 2\beta\hspace{1mm}}

\newcommand{\gev}{\,\, \mathrm{GeV}}

\newcommand{\BC}{\begin{center}}
\newcommand{\EC}{\end{center}}
\newcommand{\BE}{\begin{equation}}
\newcommand{\EE}{\end{equation}}
\newcommand{\BEA}{\begin{eqnarray}}
\newcommand{\BEAnn}{\begin{eqnarray*}}
\newcommand{\EEA}{\end{eqnarray}}
\newcommand{\EEAnn}{\end{eqnarray*}}
\newcommand{\non}{\nonumber}
\newcommand{\id}{{\rm 1\kern-.12em
\rule{0.3pt}{1.5ex}\raisebox{0.0ex}{\rule{0.1em}{0.3pt}}}}
\newcommand{\lsim}
{\;\raisebox{-.3em}{$\stackrel{\displaystyle <}{\sim}$}\;}
\newcommand{\gsim}
{\;\raisebox{-.3em}{$\stackrel{\displaystyle >}{\sim}$}\;}

\newcommand{\gf}{G_F}
\def\al{\alpha}
\def\als{\alpha_s}

\def\be{\beta}

\def\De{\Delta}

\def\hSi{\hat{\Sigma}}

\def\hSie{\hat{\Sigma}^{(1)}}
\def\hSiz{\hat{\Sigma}^{(2)}}

\newcommand{\sqmsmt}{\sqrt{\ms^2-\mt^2}}

\newcommand{\lmtms}{\KL\frac{\mt^2}{\ms^2}\KR}
\newcommand{\lmtmsms}{\KL\frac{\mtms^2}{\ms^2}\KR}

\def\draftdate{\relax}
\def\mda{\relax}
\def\mua{\relax}
\def\mla{\relax}
\def\draft{
\def\thtystars{******************************}
\def\sixtystars{\thtystars\thtystars}
\typeout{}
\typeout{\sixtystars**}
\typeout{* Draft mode!
         For final version remove \protect\draft\space in source file *}
\typeout{\sixtystars**}
\typeout{}
\def\draftdate{\today}
\def\mua{\marginpar[\boldmath\hfil$\uparrow$]%
                   {\boldmath$\uparrow$\hfil}%
                    \typeout{marginpar: $\uparrow$}\ignorespaces}
\def\mda{\marginpar[\boldmath\hfil$\downarrow$]%
                   {\boldmath$\downarrow$\hfil}%
                    \typeout{marginpar: $\downarrow$}\ignorespaces}
\def\mla{\marginpar[\boldmath\hfil$\rightarrow$]%
                   {\boldmath$\leftarrow $\hfil}%
                    \typeout{marginpar: $\leftrightarrow$}\ignorespaces}
\def\Mua{\marginpar[\boldmath\hfil$\Uparrow$]%
                   {\boldmath$\Uparrow$\hfil}%
                    \typeout{marginpar: $\Uparrow$}\ignorespaces}
\def\Mda{\marginpar[\boldmath\hfil$\Downarrow$]%
                   {\boldmath$\Downarrow$\hfil}%
                    \typeout{marginpar: $\Downarrow$}\ignorespaces}
\def\Mla{\marginpar[\boldmath\hfil$\Rightarrow$]%
                   {\boldmath$\Leftarrow $\hfil}%
                    \typeout{marginpar: $\Leftrightarrow$}\ignorespaces}
\overfullrule 5pt
\oddsidemargin -15mm
\evensidemargin -15mm
\marginparwidth 29mm
}

\begin{document}
\thispagestyle{empty}

\def\thefootnote{\fnsymbol{footnote}}

\begin{flushright}
CERN-TH/99-74\\
DESY 99-012\\
KA-TP-1-1999\\
hep-ph/9903404 \\
\end{flushright}

\vspace{1cm}

\begin{center}

{\large\sc {\bf The Mass of the Lightest MSSM Higgs Boson:}}

\vspace*{0.4cm} 

{\large\sc {\bf A Compact Analytical Expression at the Two-Loop Level}}

\vspace{1cm}

{\sc 
S.~Heinemeyer$^{1}$%
\footnote{email: Sven.Heinemeyer@desy.de}%
, W.~Hollik$^{2,3}$%
\footnote{email: Wolfgang.Hollik@physik.uni-karlsruhe.de}%
, G.~Weiglein$^{3}$%
\footnote{email: georg@particle.physik.uni-karlsruhe.de}
}

\vspace*{1cm}

{\sl
$^1$ DESY Theorie, Notkestr. 85, 22603 Hamburg, Germany

\vspace*{0.4cm}

$^2$ Theoretical Physics Division, CERN, CH-1211 Geneva 23, Switzerland

\vspace*{0.4cm}

$^3$ Institut f\"ur Theoretische Physik, Universit\"at Karlsruhe, \\
D--76128 Karlsruhe, Germany
}

\end{center}

\vspace*{1cm}

\begin{abstract}
A compact approximation formula for the mass of the lightest neutral
$\cp$-even Higgs boson, $\mh$, in the Minimal Supersymmetric Standard
Model (MSSM) is derived from the diagrammatic \twol\ result for $\mh$ up 
to $\oaas$. By analytically expanding the
diagrammatic result the leading logarithmic and non-logarithmic as well
as the dominant subleading contributions are obtained. The approximation
formula is valid for general mixing in the scalar top sector and 
arbitrary choices of the parameters in the Higgs sector of the model.
Its quality is analyzed by comparing it with the full diagrammatic
result. We find agreement with the full result better than $2 \gev$ for
most parts of the MSSM parameter space.
\end{abstract}

\def\thefootnote{\arabic{footnote}}
\setcounter{page}{0}
\setcounter{footnote}{0}

\newpage


\section{Introduction}

One of the most striking phenomenological implications of Supersymmetry
(SUSY) is the prediction of a relatively light Higgs boson, which is 
common to all supersymmetric models whose couplings remain in the
perturbative regime up to a very high energy
scale~\cite{susylighthiggs}. The search for the lightest Higgs boson thus 
allows a crucial test of SUSY, and is one of the main goals at the
present and the next generation of colliders. 
A precise knowledge of the dependence of the mass $\mh$ of the lightest
Higgs boson on the relevant SUSY parameters is necessary for a
detailed analysis of SUSY phenomenology at LEP2, the upgraded Tevatron,
and also for the LHC and a future linear $e^+e^-$ collider, where
a high-precision measurement of $\mh$ might become possible.

In the Minimal Supersymmetric Standard Model (MSSM)~\cite{mssm}, at
the tree level the mass of the lightest Higgs boson is restricted 
to be smaller than the $Z$-boson mass $\MZ$. This bound, however, is strongly 
affected by radiative corrections, 
resulting in an upper bound of about $135 \gev$~%
\cite{mhiggs1l,mhiggs1lrest,mhiggsf1l,mhiggsf1lb,mhiggsf1lc,mhiggsletter,mhiggsletter2,mhiggslong,mhiggsRG1,mhiggsRG1a,mhiggsRG2,mhiggsEffPot,mhiggsEffPot2}.
Results beyond \onel\ order have been obtained using several different
approaches: 
a Feynman-diagrammatic calculation of the leading QCD
corrections has been performed~\cite{mhiggsletter,mhiggsletter2,mhiggslong};
renormalization group (RG)
methods have been applied in order to obtain leading logarithmic
higher-order contributions~\cite{mhiggsRG1,mhiggsRG1a,mhiggsRG2};
the leading \twol\ QCD corrections have been calculated in the
effective potential method~\cite{mhiggsEffPot,mhiggsEffPot2}.
%
Until recently phenomenological analyses have been based either on
RG improved \onel\ calculations~\cite{mhiggsRG1,mhiggsRG1a,mhiggsRG2} or
on the complete Feynman-diagrammatic \onel\ on-shell
result~\cite{mhiggsf1l,mhiggsf1lb,mhiggsf1lc}. Their numerical results,
however, differ by up to $20 \gev$. 
Recently the Feynman-diagrammatic result
for the dominant contributions in $\oaas$ to the masses of the neutral
$\cp$-even Higgs bosons has become
available~\cite{mhiggsletter}. By combining these contributions with the 
complete \onel\ on-shell result~\cite{mhiggsf1lb}, the presently
most precise result for $\mh$ based on diagrammatic calculations is
obtained~\cite{mhiggsletter2,mhiggslong}. In comparison with the results
obtained by RG methods good agreement is found in the case of vanishing
mixing in the scalar top sector, while sizeable deviations which can 
exceed $5 \gev$ occur when mixing in the $\Stop$-sector is taken into
account~\cite{mhiggsletter2,mhiggslong}. 

The Feynman-diagrammatic \twol\ result for $\mh$, 
however, is very lengthy, making the 
evaluation of the Higgs-boson masses in this approach relatively slow.
This could limit the applicability of this result e.g.\ in Monte
Carlo simulations. In the present paper we derive, by means of
a Taylor expansion, a short analytical approximation formula from the 
diagrammatic \twol\ result up to $\oaas$~\cite{mhiggsletter,mhiggsletter2}.
The purpose of this is not only to provide a compact analytical expression
for $\mh$ suitable for a very fast numerical evaluation 
without losing too much of
accuracy, but also to isolate the most important contributions, thus 
allowing a better qualitative understanding of the source of the dominant 
corrections. The compact approximation formula contains, besides
the relevant parts of the \onel\ contributions, partly taken over from
\citere{mhiggsRG2}, the leading logarithmic and non-logarithmic \twol\ 
corrections for general mixing in the $\Stop$-sector, the leading Yukawa 
corrections~\cite{mhiggsRG1a,ccpw} and leading QCD contributions
beyond $\oaas$.

The values for $\mh$ obtained from the approximation formula are compared
with the full result~\cite{mhiggsletter2, mhiggslong}.
The dependence on the various MSSM parameters from the stop
sector, the Higgs sector and the chargino-neutralino sector is
analyzed. We find that the approximation formula agrees with the full 
result better than $2 \gev$ for most parts of the MSSM parameter space.

The paper is organized as follows: In Sect.~2 a Taylor expansion of
the diagrammatic \twol\ result up to $\oaas$ is performed and a compact
approximation formula is derived.
The accuracy of the approximation formula is discussed in 
Sect.~3, where also the location of the maximal values of $\mh$, depending
on the mixing in the $\Stop$-sector, is analyzed analytically. In
Sect.~4 we give our conclusions. 


\section{The compact analytical formula}

\subsection{The scalar top sector of the MSSM}

In order to fix our notations and to explain the approximations employed
in this sector, we first list the conventions for the
MSSM scalar top sector:
the mass matrix in the basis of the current eigenstates $\StopL$ and
$\StopR$ is given by
\BE
\label{stopmassmatrixwithdt}
{\cal M}^2_{\Stop} =
  \ML \MstL^2 + \mt^2 + \CZb (\edz - \frac{2}{3} \sw^2) \MZ^2 &
      \mt \Mtlr \\
      \mt \Mtlr &
      \MstR^2 + \mt^2 + \frac{2}{3} \CZb \sw^2 \MZ^2 
  \MR,
\EE
where 
\BE
\mt \Mtlr = \mt (A_t - \mu \CTb)~.
\label{eq:mtlr}
\EE
Neglecting the numerically small contributions proportional to $\MZ^2$ and
setting%
\footnote{
Later we will also discuss the case $\MstL \neq \MstR$, for which the
same formalism applies, but with a different definition for $\ms$,
see \refeq{msfkt}.
} 
\BE
\MstL = \MstR := \msq,\;\;\ms^2 := \msq^2 + \mt^2
\EE
one arives at
\BE
\label{stopmassmatrix}
{\cal M}^2_{\Stop} =
  \ML \ms^2  &  \mt \Mtlr \\
      \mt \Mtlr & \ms^2 
  \MR~.
\EE

Diagonalizing the $\Stop$-mass matrix (\ref{stopmassmatrix}) yields
the mass eigenvalues  
$\mste, \mstz$ and the $\Stop$-mixing angle~$\tst$, which relates
the current eigenstates to the mass eigenstates:
\BE
\VL \Stope \\ \Stopz \VR = \ML \costt & \sintt \\ -\sintt & \costt \MR 
                       \VL \StopL \\ \StopR \VR~.
\label{sfermrotation}
\EE

In the above approximation, which we will use throughout the rest of
the paper, the $\Stop$-masses and the mixing angle are given by
\BEA
\label{mstopins}
\mste^2 &=& \ms^2 - |\mt \Mtlr| = \ms^2 \KL 1 - \dst \KR , \non \\
\mstz^2 &=& \ms^2 + |\mt \Mtlr| = \ms^2 \KL 1 + \dst \KR , \\
\label{mixangins}
\tst &=& \KKKL \renewcommand{\arraystretch}{1.3}
         \begin{array}{r@{~~~}l}
               \frac{\pi}{4} & \mbox{for } \Mtlr < 0 \\
               0 & \mbox{for } \Mtlr = 0 \\
               -\frac{\pi}{4} & \mbox{for } \Mtlr > 0
         \end{array} \right. , ~
         \renewcommand{\arraystretch}{1}
\EEA
with
\BE
\dst = \frac{|\mt \Mtlr|}{\ms^2} = 
       \frac{\mstz^2 - \mste^2}{\mstz^2 + \mste^2}~.
\EE 
{}From the above definition it follows that $0 \le \dst \le 1$, otherwise
the $\Stop$-mass matrix \refeq{stopmassmatrix} would have a negative
eigenvalue. 


\subsection{Calculation of the mass of the lightest Higgs boson}
\label{subsec:mhcalc}

Here we only give a very brief outline of the calculation of the mass of
the lightest neutral $\cp$-even Higgs boson in the MSSM;
for notations and a detailed description see
\citeres{mhiggsletter,mhiggsletter2,mhiggslong}. 
We focus on the different steps of approximations made in order
to derive a compact analytical expression from the full diagrammatic
result of $\oaas$.


\bigskip
At the tree level, 
the mass matrix of the neutral $\cp$-even Higgs bosons
in the basis of the weak eigenstates $\Pe,\Pz$ 
can be expressed 
in terms of $\MZ$, $\MA$ (the mass of the $\cp$-odd Higgs boson) and
$\tb = v_2/v_1$ (the ratio of the vacuum expectation values of the two
Higgs doublets) as follows:
\BEA
\label{higgsmasstree}
M_{\rm Higgs}^{2, {\rm tree}} &=& \ML \mpe^2 & \mpez^2 \\ 
                           \mpez^2 & \mpz^2 \MR \non\\
&=& \ML \MA^2 \SQb + \MZ^2 \CQb & -(\MA^2 + \MZ^2) \Sb \Cb \\
    -(\MA^2 + \MZ^2) \Sb \Cb & \MA^2 \CQb + \MZ^2 \SQb \MR .
\EEA

Taking into account higher-order corrections, the Higgs-boson masses
can to a good approximation be obtained by diagonalizing the matrix
\BE
M^{2}_{\rm Higgs}
= \VL \mpe^2 - \hSi_{\Pe}(0)\;\;\;\;\;\; \mpez^2 - \hSi_{\PePz}(0) \\
     \mpez^2 - \hSi_{\PePz}(0)\;\;\;\;\;\; \mpz^2 - \hSi_{\Pz}(0) \VR~,
\label{higgsmassmatrixnondiag}
\EE
where the $\hSi_s(0)~(s = \Pe, \PePz, \Pz)$ denote the
renormalized Higgs-boson self-energies (in the $\Pe,\Pz$ basis). 
As a first step of approximation, the momentum dependence, which is 
numerically rather small, has been neglected in the $\hSi_s(p^2)$.

\bigskip
The mass of the lightest Higgs boson receives contributions from all
sectors of the MSSM, but not all are numerically of equal relevance.
In order to derive a compact analytical expression, we make the following 
further approximations:


\begin{description}

\item[The contributions from the $t,\Stop$-sector up to the \twol\ level:]
\label{subsubsec:mh2top} 
The contributions\\
arising from the $t,\Stop$-sector can be written as
\BE
\hSi_s(0) = \hSie_s(0) + \hSiz_s(0),~s = \Pe, \PePz, \Pz .
\EE
Here the $\hSie_s(0)$ denote the \onel\ contributions of the
$t,\Stop$-sector to the renormalized Higgs-boson self-energies. Their
explicit form (including also the momentum dependence) can be found e.g.\
in \citere{mhiggsf1lb}. The $\hSiz_s(0)$ denote the \twol\ contributions
from the $t,\Stop$-sector at zero external momentum from the Yukawa part
of the theory (neglecting the gauge couplings)~\cite{mhiggsletter}.
We first consider the dependence of the mass shift $\De\mh^{2}$ ($\mh^2
= \mh^{2,{\rm tree}} + \De\mh^{2}$) on $\tb$ and $\mu$. For the leading
contributions to $\De\mh^{2}$ from the Yukawa part of the theory the
dependence on $\tb$ and $\mu$ drops out in the limit $M_A \gg M_Z$.%
\footnote{Note that $\Mtlr$ is treated as a free parameter in \refeq{eq:mtlr}
and therefore does not depend on $\mu$.}
This holds both for the contributions in $\oa$ and $\oaas$.
Since $\tb$ and $\mu$ enter only via non-leading corrections in
$\De\mh^{2}$, the dependence on them is relatively mild. We therefore
use the approximation $\mu = 0$ in the $\hSi_s(0)$. We furthermore
extract a common prefactor $(1/\sin^2\beta)$ and set otherwise 
$\sbe = 1$ in the non-logarithmic \onel\ contributions, while the full
dependence on $\sbe$ is kept in the logarithmic \onel\ and the 
\twol\ contributions.
Since the variation of $\mh$ with $\mgl$ is $\pm 2 \gev$ at most, see
\citere{mhiggslong}, we have eliminated the dependence of the
$\hSiz_s(0)$ on the gluino mass by setting
\BE
\mgl = \msusy \equiv \sqmsmt ,
\label{eq:msusymgl}
\EE
where in the case $\MstL = \MstR = \msq$ the SUSY scale is given by 
$\msusy = \msq$.

As the main step of our approximations, we have performed a Taylor
expansion in $\dst$ of the $\hSi_s(0)$ by inserting
\refeq{mstopins} for the $\Stop$-masses and \refeq{mixangins} for the
$\Stop$-mixing angle.
For the \onel\ correction we have expanded up to 
${\cal O}(\dst^8)$; 
all three renormalized \onel\ Higgs-boson self-energies give a contribution.
In the \onel\ self-energies we have kept terms up to
${\cal O}(\MZ^4/\mt^4)$ while terms of ${\cal O}(\MZ^2/\ms^2)$ have
been neglected. We have checked that the numerical effect of the
latter terms is insignificant.
Concerning the \twol\ self-energies,
the expansion has been carried out
up to ${\cal O}(\dst^4)$. 
With the above described approximations 
only $\hSiz_{\Pz}$ gives a non-zero contribution.
The Taylor expanded self-energies have then been inserted into the
Higgs-boson mass matrix \refeq{higgsmassmatrixnondiag}.


\item[The \onel\ contributions from the other sectors:]
\label{subsubsec:mh2onelooprest}

For the \onel\ corrections from the other sectors of the MSSM we use
the logarithmic approximation given in \citere{mhiggs1lrest}. In this 
approximation the scale of the soft SUSY-breaking parameter in the
gaugino sector, $M$, is chosen as $M = \msusy$, where $\msusy$ is
defined as in \refeq{eq:msusymgl}. Besides $\msusy$, this
contribution is parameterized in terms of $\MA$, the mixing angle $\be$,
and the SM parameters, see \refeq{mh2onelooprest}.
A higher accuracy of the non-leading 
\onel\ contributions in the approximation
formula can be achieved by including further terms of the \onel\
logarithmic approximation given in \citere{mhiggsRG2}.


\item[Corrections beyond $\oaas$:]
\label{subsubsec:beyondoaas}

Leading contributions beyond $\oaas$ have been taken into account by
incorporating the leading \twol\ Yukawa 
correction of ${\cal O}(\gf^2\mt^6)$ \cite{mhiggsRG1a,ccpw} and by 
expressing the $t,\Stop$-contributions
through the \msbar\ top-quark mass
\BE
\mtms = \mtms(\mt) \approx \frac{\mt}{1 + \frac{4}{3\,\pi} \als(\mt)}~
\label{mtrun}
\EE
instead of the pole mass $\mt$.
This leads to an additional contribution in 
${\cal O}(\al\als^2)$.
\end{description}


\subsection{The analytical approximation formula for $\mh$}

\subsubsection{The case for general $\MA$}
\label{subsubsec:generalMA}

With the approximations described above we obtain the following
contributions to the renormalized Higgs-boson self-energies 
from the $t,\Stop$-sector (expressed in terms of the top-quark pole
mass) at \onel\ order:
\BEA
\label{p1seol}
\hSie_{\Pe}(0) &=&
    \frac{\gf\wz}{\pi^2} \MZ^4 \Lambda \CQb
    \log\lmtms~, \\
\label{p1p2seol}
\hSie_{\PePz}(0) &=&
    -\frac{\gf\wz}{\pi^2} \MZ^2 \CTb 
    \KKL -\frac{3}{8} \mt^2 + \MZ^2 \Lambda \SQb \KKR \log\lmtms~, \\
\label{p2seol}
\hSie_{\Pz}(0) &=&
    \frac{\gf\wz}{\pi^2} \frac{\mt^4}{8 \SQb} 
      \Bigg\{ -2 \frac{\MZ^2}{\mt^2} 
              + \frac{11}{10}\frac{\MZ^4}{\mt^4} \non \\
 && {} + \KKL 12 - 6 \frac{\MZ^2}{\mt^2} \SQb +
                 8 \frac{\MZ^4}{\mt^4} \Lambda \SVb \KKR \log\lmtms \non \\
 && {} + \frac{\Mtlrz}{\ms^2}\, 
      \KKL -12 + 4 \frac{\MZ^2}{\mt^2}
          + 6 \frac{\mt^2}{\ms^2} \KKR 
    + \frac{\Mtlrv}{\ms^4}\,
      \KKL 1 - 4\,\frac{\mt^2}{\ms^2} 
           +3 \frac{\mt^4}{\ms^4} \KKR \non \\
 && {} + \frac{\Mtlrse}{\ms^6}\,
      \KKL  \frac{3}{5}\frac{\mt^2}{\ms^2}
           -\frac{12}{5}\frac{\mt^4}{\ms^4}
           +2\frac{\mt^6}{\ms^6} \KKR \non \\
 && {} + \frac{\Mtlra}{\ms^8}\,
      \Bigg[  \frac{3}{7}\frac{\mt^4}{\ms^4}
             -\frac{12}{7}\frac{\mt^6}{\ms^6}
             +\frac{3}{2}\frac{\mt^8}{\ms^8} \Bigg]
    \Bigg\}~, 
\EEA
with
\BE
\Lambda = \left(\frac{1}{8} - \frac{1}{3} \sw^2 + \frac{4}{9} 
                 \sw^4 \right), \qquad \sw^2 = 1 - \frac{\MW^2}{\MZ^2} . 
\label{eq:lambda}
\EE
We have verified that the logarithmic terms in \refeqs{p1seol}--(\ref{p2seol}) 
agree with the ones given in \citere{mhiggs1lrest}. 

The \twol\ contributions read:
\BEA
\hSiz_{\Pe}(0) &=& 0~, \non \\
\hSiz_{\PePz}(0) &=& 0~, \non \\
\label{p2setl}
\hSiz_{\Pz}(0) &=&
    \frac{\gf\wz}{\pi^2} \frac{\als}{\pi} \frac{\mt^4}{\SQb}
      \Bigg[ 3 \log^2\lmtms -6 \log\lmtms 
             -6 \frac{\Mtlr}{\ms} \non \\
 &&  {} \hspace{8em} - 3 \frac{\Mtlrz}{\ms^2} \log\lmtms  
             +\frac{3}{4} \frac{\Mtlrv}{\ms^4} \Bigg]~.
\EEA
$\ms$ has to be chosen according to
\BE
\label{msfkt}
\ms = \KKKL \begin{array}{l@{\quad:\quad}l}
            \sqrt{\msq^2 + \mt^2} & \MstL = \MstR = \msq \\
            \KKL \MstL^2 \MstR^2 + \mt^2 (\MstL^2 + \MstR^2) +
                     \mt^4 \KKR^\frac{1}{4} &
                                    \MstL \neq \MstR
            \end{array} \right. 
\EE
The last formula requires some explanation: when performing the
expansion in $\dst$, 
it was assumed that $\MstL = \MstR$. Thus our
result contains only one soft SUSY-breaking scale $\ms$. For the case
$\MstL \neq \MstR$, $\ms$ is chosen to reproduce the argument of the
leading log correctly. According to \refeq{msfkt} the log yields in
both cases 
\BE
\log\KL \frac{\mt^2}{\ms^2} \KR = 
\log\KL \frac{\mt^2}{\mste\,\mstz} \KR + {\cal O}\KL \dst \KR
\EE 
in agreement with \refeq{mstopins}.

\smallskip
The \onel\ contributions from the other sectors of the MSSM are not
listed here but can be
found in \citeres{mhiggs1lrest,mhiggsRG2}. The combined self-energies
have to be inserted into \refeq{higgsmassmatrixnondiag}. 
In order to incorporate the leading QCD corrections beyond $\oaas$, the
top-quark pole mass should be replaced by the \msbar\ top-quark mass
\refeq{mtrun} in the \twol\ contribution, as described
above. Diagonalization of \refeq{higgsmassmatrixnondiag} 
yields the square of the masses of the neutral $\cp$-even Higgs
bosons. In order to incorporate leading contributions in 
${\cal O}(\al^2)$, the leading Yukawa correction, see \refeq{mh2yuk} below,
can be added.


\subsubsection{The case $\MA \gg \MZ$}
\label{subsubsec:largeMA}

The diagonalization of the mass matrix \refeq{higgsmassmatrixnondiag}
in the evaluation of $\mh^2$ incorporates contributions that are formally 
of higher order but are non-negligible in general.
For large $\MA$ these higher-order contributions are
suppressed by inverse powers of $\MA$. Therefore it is possible 
in this case to
perform an expansion in the loop order, leading to a very compact
formula for $\mh^2$ of the form
\BE
\label{mh2lle}
\mh^2 = \mh^{2,{\rm tree}} + \De\mh^{2,\al,t/\Stop} 
         + \De\mh^{2,\al,{\rm rest}}
         + \De\mh^{2,\al\als} + \Delta\mh^{2,\al^2} .
\EE

At the \twol\ level,  only the $\Pz$ self-energy
contributes in our approximation, yielding the term
\BE
\De\mh^{2,\al\als} = - \SQb\,\hSiz_{\Pz}(0)~
\label{delmh2loop}
\EE
with $\hSiz_{\Pz}(0)$ from \refeq{p2setl}.
In contrast to \refeqs{p1seol}--(\ref{p2setl}) we give here an
expression for $\mh$ in which the \msbar\ top-quark mass is used
everywhere. Inserting \refeq{mtrun} into 
$\hSie_{\Pz}(0)$ yields additional contributions to
$\De\mh^{2,\al\als}$, see \refeq{mh2twolooptop}. 
{}From \refeq{delmh2loop} it becomes also clear that the
$\be$~dependence of the leading \twol\ contribution drops out, as
mentioned above.

\smallskip
\noindent
The tree-level and \onel\ contributions of \refeq{mh2lle} are given by
\BEA
\label{mh2tree}
\mh^{2,{\rm tree}} &=& \edz \KKL \MA^2 + \MZ^2 
         - \sqrt{(\MA^2 + \MZ^2)^2 - 4 \MZ^2 \MA^2 \CQZb} \KKR , \\
\label{mh2onelooptop}
\lefteqn{ \hspace{-3em}
\De\mh^{2,\al,t/\Stop} = \frac{\gf\wz}{\pi^2}\; \mtms^4
      \Biggl[ \log\lmtmsms 
             \Biggl\{ - \frac{3}{2} 
                     - \frac{3}{4} \frac{\MZ^2}{\mtms^2} \CZb 
                     - \frac{\MZ^4}{\mtms^4} \Lambda \CQZb \non }\\
 && {}               - \frac{\MZ^2}{\MA^2} \CQb \CZb
                       \left( 6 + \frac{3}{2} \frac{\MZ^2}{\mtms^2} 
                           (1 - 4 \SQb) 
                     - \frac{\MZ^4}{\mtms^4} 8 \Lambda \CZb \SQb 
                       \right) \Biggr\} \non \\
 && {}         +\Bigg\{  \frac{1}{4} \frac{\MZ^2}{\mtms^2}
                     -\frac{11}{80} \frac{\MZ^4}{\mtms^4}  
           + \frac{\Mtlrz}{\ms^2} 
                    \KL \frac{3}{2} - \edz \frac{\MZ^2}{\mtms^2}
                       - \frac{3}{4} \frac{\mtms^2}{\ms^2} \KR \non\\
 && {}               +\frac{\Mtlrv}{\ms^4}
                    \KL -\frac{1}{8} + \edz \frac{\mtms^2}{\ms^2} 
                        -\frac{3}{8} \frac{\mtms^4}{\ms^4}
                    \KR 
                  +\frac{\Mtlrse}{\ms^6}
                    \KL -\frac{3}{40} \frac{\mtms^2}{\ms^2}
                        +\frac{3}{10} \frac{\mtms^4}{\ms^4} 
                        -\frac{1}{4} \frac{\mtms^6}{\ms^6} \KR \non\\
 && {}               +\frac{\Mtlra}{\ms^8}
                    \KL -\frac{3}{56}\frac{\mtms^4}{\ms^4} 
                        +\frac{3}{14}\frac{\mtms^6}{\ms^6}
                        -\frac{3}{16}\frac{\mtms^8}{\ms^8} \KR
   \Bigg\} \KL 1 + 4 \frac{\MZ^2}{\MA^2} \CQb \CZb \KR \Biggr] , \\
\label{mh2onelooprest}
\lefteqn{ \hspace{-3em}
\De\mh^{2,\al,{\rm rest}} = \frac{\gf\wz}{\pi^2} \frac{\MZ^4}{24} 
             \Biggl[ \log\KL\frac{\msusy^2}{\MZ^2}\KR \Bigg\{
                12 N_c \frac{\mb^4}{\MZ^4} 
                - 6 N_c \CZb \frac{\mb^2}{\MZ^2} + \CQZb (P_b + P_f) \non } \\
 && {}             + (P_g + P_{2H}) (\SVb + \CVb)
                - 2 \SQb \CQb (P'_g + P'_{2H}) \Bigg\} 
                +\frac{6 N_c \mb^4}{\MZ^4} \times \non \\
 && {}        \Bigg[ 2 \frac{\Mblrz}{\msusy^2} 
                   \KL 1 - \frac{\Mblrz}{12\,\msusy^2} \KR
                   - \frac{\MZ^2}{2\,\mb^2}\CZb
                     \KL \frac{\Mblrz}{\msusy^2} + 
                         \frac{A_b^2 - \mu^2\TQb}{3\,\msusy^2} \KR
            \Bigg] \non \\
 && {}        - \log\KL\frac{\MA^2}{\MZ^2}\KR \Biggl\{
                (\CVb + \SVb) P_{2H} - 2 \CQb \SQb P'_{2H} - P_{1H}
                \Biggr\} \Biggr] ,
\EEA
where $\Lambda$ is defined as in \refeq{eq:lambda}, and
\BEA
\Mblr &=& A_b - \mu \tb \non \\
P_b &=& N_c (1 + 4 Q_b \sw^2 + 8 Q_b^2 \sw^4) \non \\
P_f &=& N_c (N_g - 1) (2 - 4 \sw^2 + 8 (Q_t^2 + Q_b^2) \sw^4)
        + N_g (2 - 4 \sw^2 + 8 \sw^4) \non \\
P_g &=& - 44 + 106 \sw^2 - 62 \sw^4,~~P'_g~=~10 + 34\sw^2 - 26\sw^4 \non \\
P_{1H} &=& -9 \CVZb + (1 - 2 \sw^2 + 2 \sw^4) \CQZb \non \\
P_{2H} &=& -10 + 2 \sw^2 - 2 \sw^4,~~P'_{2H}~=~8 - 22 \sw^2 + 10\sw^4 \non \\
 && Q_t = \frac{2}{3},\; Q_b = -\frac{1}{3},\;
               N_c = 3,\; N_g = 3 ~.
\EEA
The dominant \twol\ contribution of $\oaas$ to $\mh^2$ reads:
\BEA 
\label{mh2twolooptop}
\De\mh^{2,\al\als} &=&
    - \frac{\gf\wz}{\pi^2} \frac{\als}{\pi}\; \mtms^4
      \Biggl[ 4 + 3 \log^2\lmtmsms + 2 \log\lmtmsms 
             -6 \frac{\Mtlr}{\ms} \non \\
 && {}  
     - \frac{\Mtlrz}{\ms^2} \KKKL 3 \log\lmtmsms +8 \KKKR \non \\
 && {} +\frac{17}{12} \frac{\Mtlrv}{\ms^4} \Biggr] 
         \KL 1 + 4 \frac{\MZ^2}{\MA^2} \CQb \CZb \KR .
\EEA
For the leading \twol\ Yukawa correction we use the result given
in~\cite{mhiggsRG1a,ccpw}:
\BEA
\label{mh2yuk}
\Delta\mh^{2,\al^2} &=& \frac{9}{16\pi^4} G_F^2 \mtms^6
               \KKL \tilde{X} t + t^2 \KKR~, \\
\tilde{X} &=& \Bigg[
                \KL \frac{\mstz^2 - \mste^2}{4 \mtms^2} \sinQZtt \KR^2
                \KL 2 - \frac{\mstz^2 + \mste^2}{\mstz^2 - \mste^2}
                      \log\KL \frac{\mstz^2}{\mste^2} \KR \KR \non\\
            && {} 
               + \frac{\mstz^2 - \mste^2}{2 \mtms^2} \sinQZtt
                      \log\KL \frac{\mstz^2}{\mste^2} \KR \Bigg], 
              \quad t = 
              \frac{1}{2} \log \KL \frac{\mste^2 \mstz^2}{\mtms^4} \KR .
\EEA
Here $\ms$ is chosen in analogy with \refeq{msfkt}, where the
top-quark pole mass $\mt$ has to be replaced by the \msbar\ top-quark
mass $\mtms$.

In the contributions of the $t,\Stop$-sector at \onel\ and \twol\ order,
\refeqs{mh2onelooptop} and (\ref{mh2twolooptop}), we have included
correction factors of ${\cal O}(\MZ^2/\MA^2)$. In this way the compact
formula \refeq{mh2lle} gives a reliable approximation for $\MA$ values
down to at least $\MA = 200 \gev$.

\bigskip
The contribution of $\oaas$ given in \refeq{mh2twolooptop} can 
be compared with analytical formulas derived via the \twol\ effective
potential approach for the case of no mixing in the
$\Stop$-sector~\cite{mhiggsEffPot}
and via RG methods~\cite{mhiggsRG1a,mhiggsRG2}. The leading 
term $\sim\log^2(\mtms^2/\ms^2)$ agrees with the results in
\citeres{mhiggsRG1a,mhiggsRG2,mhiggsEffPot}. The subleading term for
vanishing $\Stop$-mixing $\sim\log(\mtms^2/\ms^2)$ agrees with the 
result of the \twol\ effective potential approach~\cite{mhiggsEffPot}
and the result of the \twol\ RG calculation~\cite{mhiggsEffPot,mhiggsRG2},
but differs from the RG improved \onel\
result~\cite{mhiggsRG1a,mhiggsRG2}. The term
$\sim\log(\mtms^2/\ms^2)(\Mtlr/\ms)^2$ for non-vanishing $\Stop$-mixing
differs from the result given in~\citere{mhiggsRG1a,mhiggsRG2}.
All other terms of $\oaas$ are new.
The term $\sim\Mtlr/\ms$ shows that the result for $\mh$ is not
symmetric in $\pm\Mtlr$.
The good numerical agreement with the RG results in the case of no
mixing in the $\Stop$-sector can qualitatively be understood by noting
that in the no-mixing case the leading term in both approaches agrees,
while for the corrections proportional to powers of $\Mtlr/\ms$
deviations occur already in the leading contribution. 

When comparing the results of the diagrammatic on-shell calculation
with the RG results in terms of the parameters 
$\ms$ and $\Mtlr$ it should be noted that, due to the different
renormalization schemes employed, the meaning of these
(non-observable) parameters is not precisely the same in the two
approaches starting from \twol\ order (see the discussion in
\citere{mhiggsletter2}). A more detailed comparison of our results
with those obtained via RG methods will be performed in
a forthcoming publication.


\subsection{Implementation into \fh}

The formulas given in the previous section
have been implemented into the Fortran
code \fh~\cite{feynhiggs}, thus allowing a direct comparison
between the full result described in
\citeres{mhiggsletter2,mhiggslong} and the approximation
formula~(\ref{mh2lle}).

We also provide the Fortran code \fhf\ in which only the formula for
the compact approximation of $\mh$ is implemented,
thus allowing an approximate but much
faster evaluation of the Higgs-boson mass $\mh$.
This program is shorter by a factor of 50 and faster by a factor of
$3 \times 10^4$ with respect to \fh;
it needs only $2 \times 10^{-5}$ seconds on a Sigma station 
(Alpha CPU, 600 MHz) for the
evaluation of $\mh$ for one set of parameters. 
The quality of the prediction of $\mh$ is reasonably good for a large part 
of the MSSM parameter space, as will be
discussed in the next section.
Into \fhf\ we have
also implemented the calculation of the MSSM contributions to
$\De\rho$~\cite{drhosuqcd}. Here the corrections 
arising from $\Stop/\Sbot$-loops up to $\oaas$
have been taken into account, neglegting only the gluino-exchange
contribution which is very lengthy and vanishes for large $\mgl$.
The $\rho$-parameter can be used as an additional constraint (besides
the experimental bounds) on the squark masses. 
A value of $\De\rho$ outside the experimentally preferred region of 
$\dr^{\SU} \lsim 1\times 10^{-3}$~\cite{delrhoexp} indicates experimentally
disfavored $\Stop$- and $\Sbot$-masses.

Both Fortran codes can be obtained via the WWW page\\
{\tt http://www-itp.physik.uni-karlsruhe.de/feynhiggs}~.


\section{Discussion of the compact expression for $\mh$}

In this section we analyze the quality of our
compact approximation formula for $\mh$ with respect to the full
calculation, which contains the full diagrammatic \onel\
contribution~\cite{mhiggsf1lb}, the complete leading \twol\
corrections in $\oaas$~\cite{mhiggsletter,mhiggsletter2,mhiggslong},
and the two contributions beyond $\oaas$, see \refeq{mtrun} 
and \refeq{mh2yuk}. 
We use here the approximation formula for general $\MA$, as discussed
in Sec.~\ref{subsubsec:generalMA}. As mentioned above, in the region 
$\MA \gsim 200 \gev$ the compact formula given in
Sec.~\ref{subsubsec:largeMA} yields equally good results.

For $\tb$ we restrict ourselves
to two typical values which
are favored by SUSY-GUT scenarios~\cite{su5so10}: $\tb = 1.6$ for
the $SU(5)$ scenario and $\tb = 40$ for the $SO(10)$ scenario. 
Other parameters are $\MZ = 91.187 \gev,~\MW = 80.39 \gev,~
G_F = 1.16639\times 10^{-5} \gev^{-2},~\als(\mt) = 0.1095$,~ 
$\mt = 175 \gev$, and $\mb = 4.5 \gev$. In the numerical evaluation we
have furthermore chosen $A_t = A_b$.
The parameter $M$ appearing in the plots is the $SU(2)$ gaugino mass
parameter, it enters in the full result only, see the discussion of
the \onel\ contributions in \refse{subsec:mhcalc}.
$\mu$ is the Higgs-mixing parameter.
If not indicated differently, we have chosen $M = \msq$ and 
$\mu = -\msq$.

\begin{figure}[th!]
\begin{center}
\mbox{
\psfig{figure=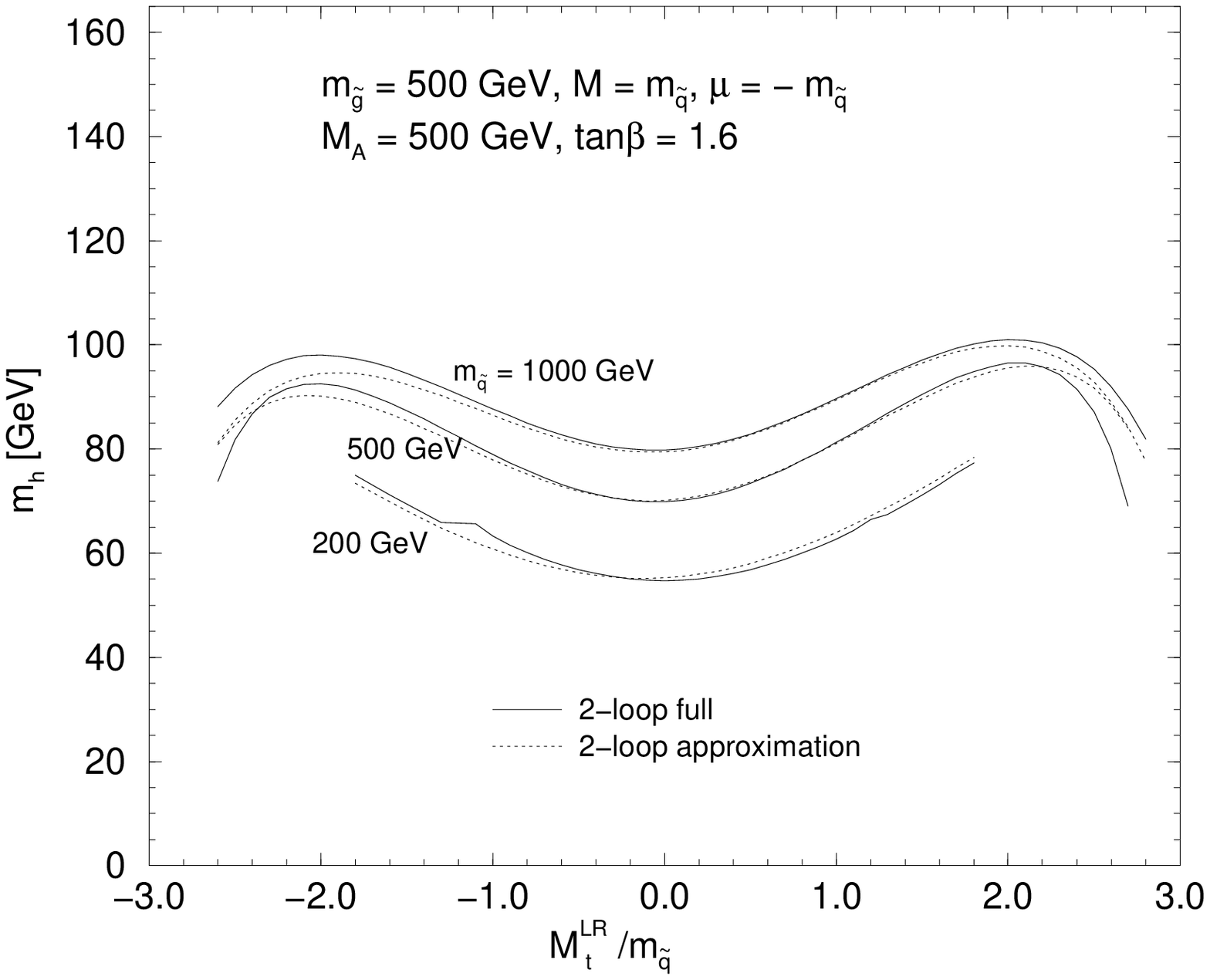,width=7cm,height=7cm}}
\hspace{1em}
\mbox{
\psfig{figure=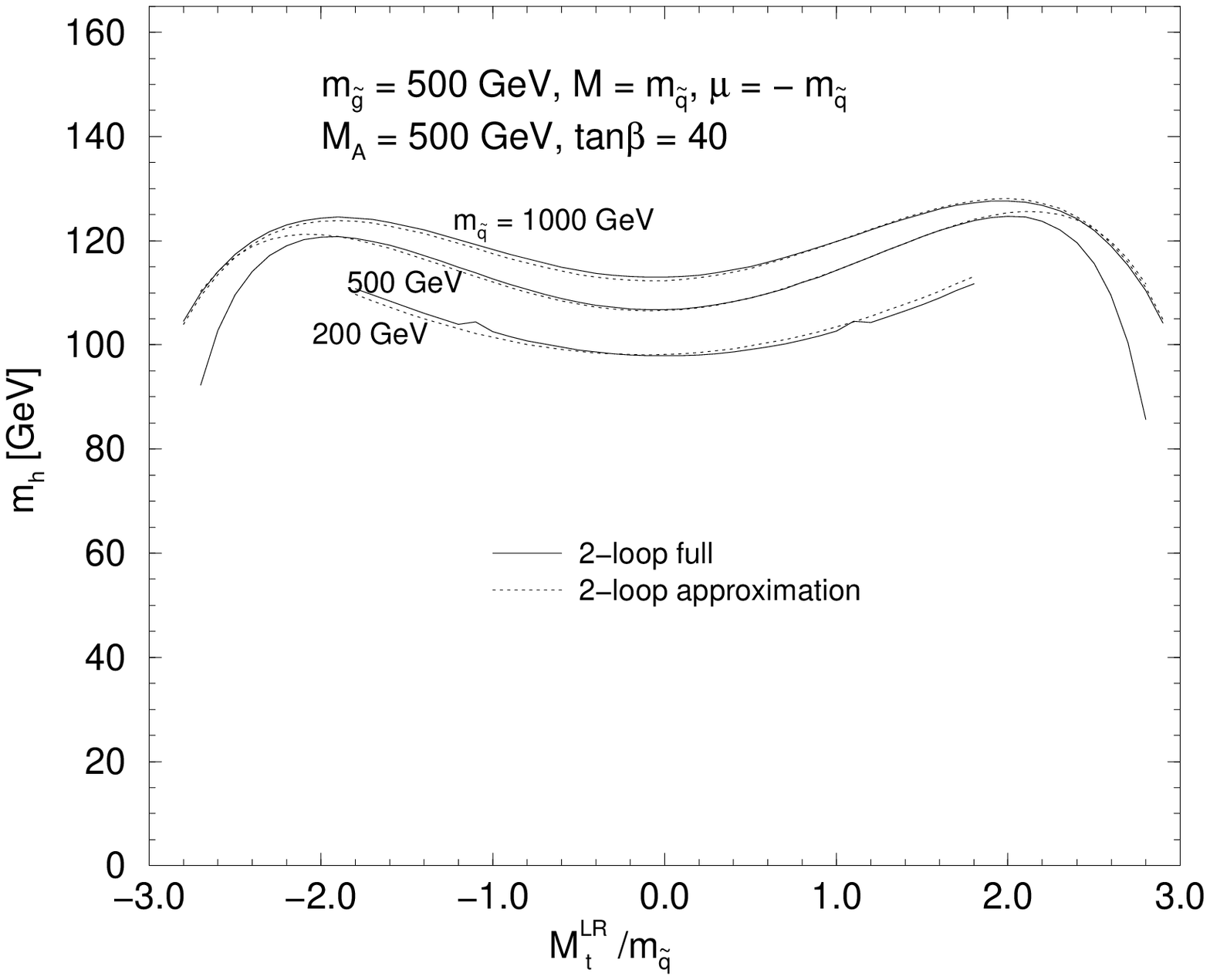,width=7cm,height=7cm}}
\end{center}
\vspace{-1.5em}
\caption[]{
$\mh$ as a function of $\Mtlr/\msq$, calculated from the full formula and
from the  approximation formula for $\MA = 500 \gev, 
\mgl = 500 \gev$ and $\tb = 1.6$ or $40$.
}
\label{fig:mh_MtLRdivmq}
\end{figure}

\bigskip
In \reffi{fig:mh_MtLRdivmq} we show $\mh$ as a function of
$\Mtlr/\msq$ for $\msq = 200, 500, 1000 \gev$ and $\MA = 500 \gev$.
A maximum for $\mh$, evaluated with the full formula as well as with
the approximation formula, is reached for about 
$\Mtlr/\msq \approx \pm 1.9$ 
in the $\tb = 1.6$ scenario and in the  
$\tb = 40$ scenario. This case we refer to as `maximal mixing'.  A
minimum is reached around $\Mtlr/\msq \approx 0$ which 
we refer to as 'no mixing'. In general the approximation differs from
the full result by less than $2 \gev$ up to $|\Mtlr/\msq| \lsim 2$.
For larger $|\Mtlr|$ sizeable deviations occur, which become very
large for $|\Mtlr/\msq| \gsim 2.5$. The reason is that the expansion
parameter~$\dst$ becomes rather large and approaches 1 in this
region. 
The effect of the non-logarithmic terms in the \twol\ contribution
in $\oaas$ reaches up to about $5 \gev$ for maximal mixing.

The location of the extrema can be understood analytically as follows:
taking into account only the leading \onel\ corrections from the 
$t,\Stop$-sector for the calculation of $\mh$, one can easily compute
the values of $\Mtlr/\msq$ for which $\mh$ reaches a maximum or a
minimum. The well known results are
\BE
\frac{\Mtlr}{\msq} = \KKKL \renewcommand{\arraystretch}{1.3}
         \begin{array}{l r}
               \sqrt{6}  & \mbox{(maximum)}\\
               0         & \mbox{(minimum)} \\
               -\sqrt{6} & \mbox{(maximum)}
         \end{array} \right. .
         \renewcommand{\arraystretch}{1}
\label{mh21loopextrema}
\EE
Taking into account the new \twol\ corrections $\De\mh^{2,\al\als}$,
given in eq.~(\ref{mh2twolooptop}), the positions of the extrema
in eq.~(\ref{mh21loopextrema}) receive a shift, yielding up to $\oaas$:
\BE
\frac{\Mtlr}{\msq} = \KKKL \renewcommand{\arraystretch}{1.3}
         \begin{array}{l l}
               \sqrt{6} -\frac{\als}{\pi}
                \KKL -1 + 3 \sqrt{6} - \sqrt{6} \log \lmtmsms \KKR 
                 & \quad (\approx +1.92 \mbox{~for $\ms = 1000 \gev$;
                   max.})\\
               -2 \frac{\als}{\pi} & 
                 \quad (\mbox{$\approx -0.07$; minimum}) \\
               -\sqrt{6} +\frac{\als}{\pi} 
                \KKL 1 + 3 \sqrt{6} - \sqrt{6} \log \lmtmsms \KKR
                 & \quad (\approx -1.85 \mbox{~for $\ms = 1000 \gev$; max.})
         \end{array} \right. ~
         \renewcommand{\arraystretch}{1}
\label{mh22loopextrema}
\EE
The maxima are shifted to smaller absolute values of $\Mtlr/\msq$, the
minimum is shifted 
to a slightly negative value (see the discussion in
\citeres{mhiggsletter, mhiggsletter2, mhiggslong}).

\begin{figure}[ht!]
\begin{center}
\mbox{
\psfig{figure=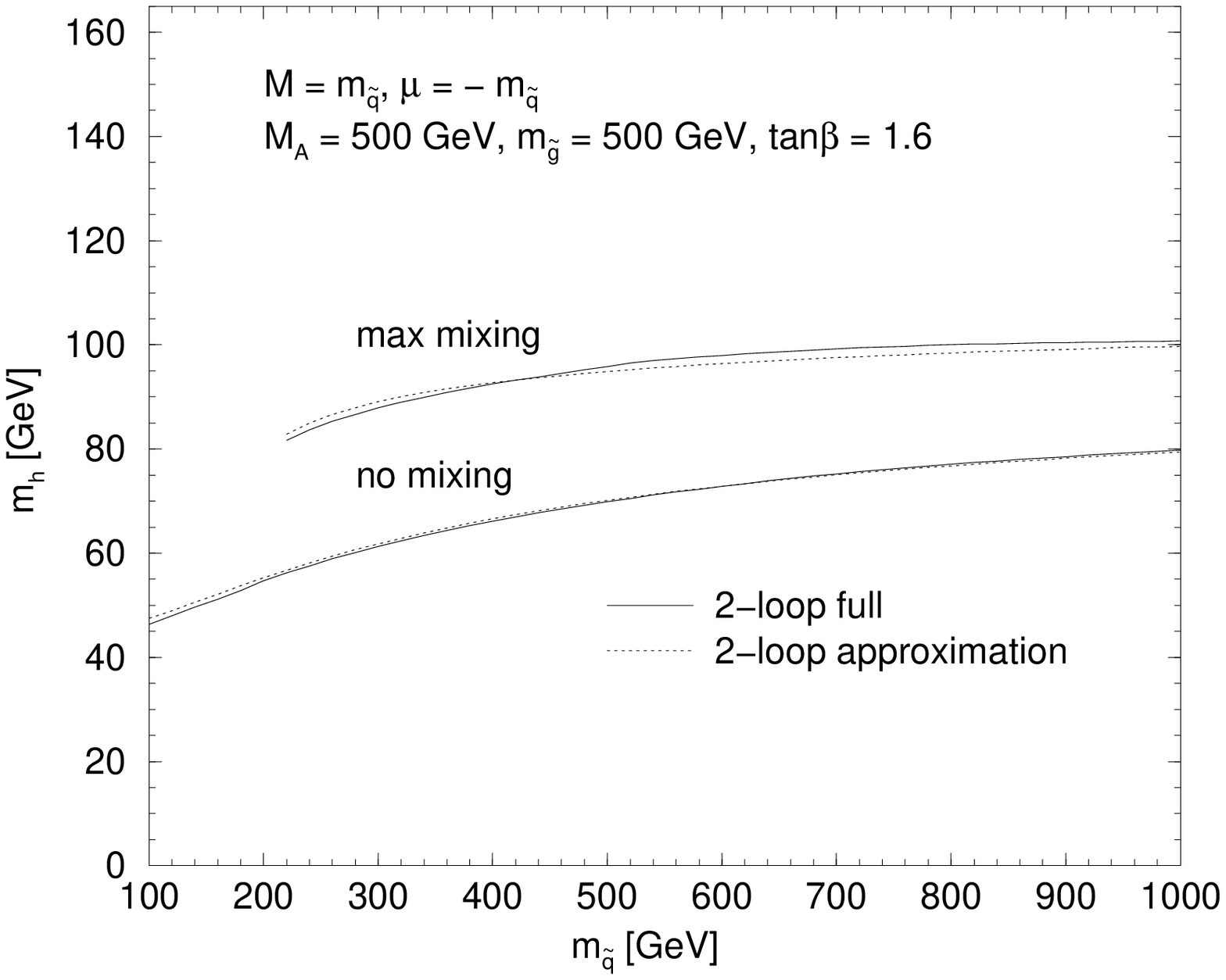,width=7cm,height=7cm}}
\hspace{1em}
\mbox{
\psfig{figure=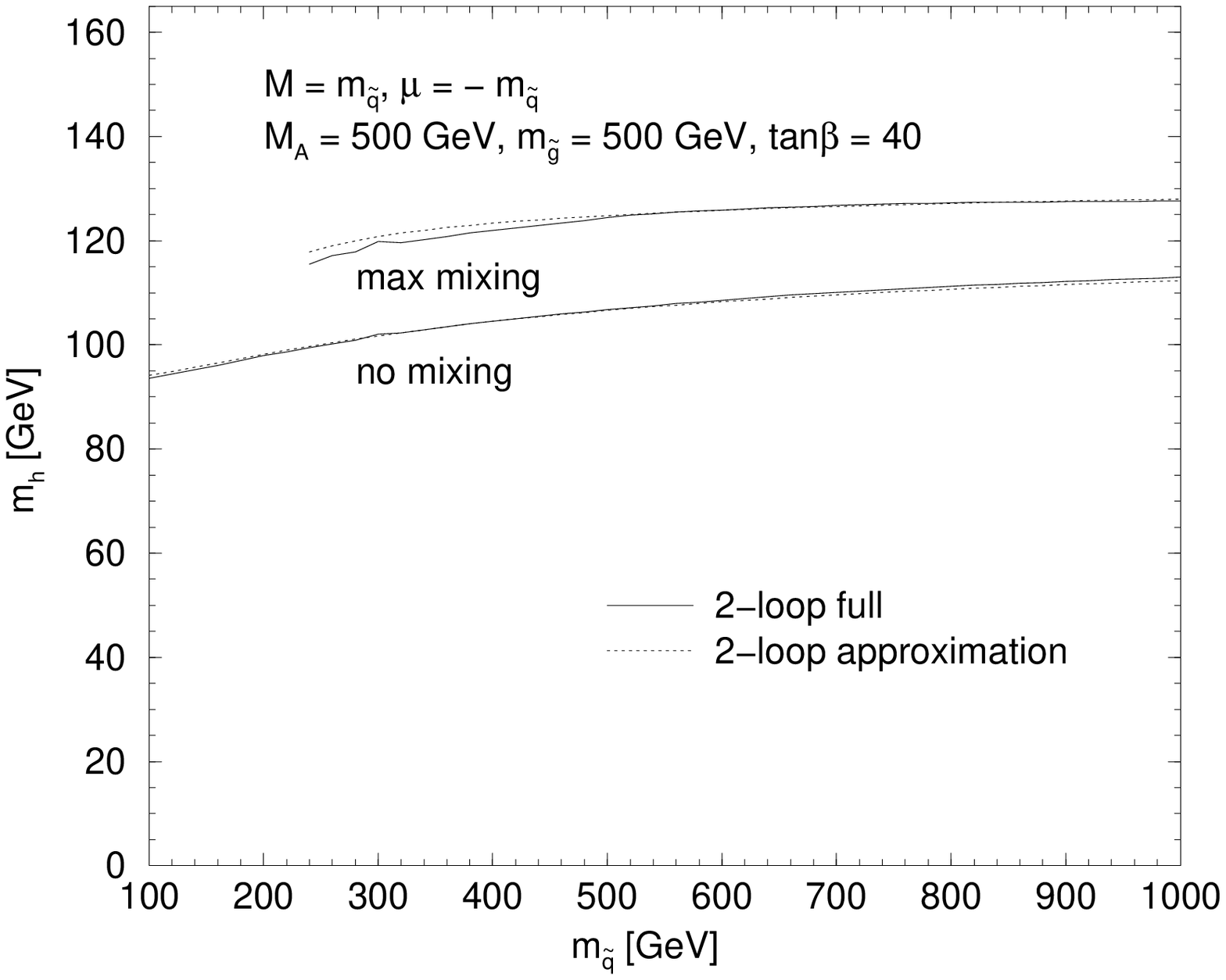,width=7cm,height=7cm}}
\end{center}
\vspace{-1.5em}
\caption[]{
$\mh$ as a function of $\msq$, calculated from the full formula and
from the approximation formula for $\MA = 500 \gev, 
\mgl = 500 \gev$ and $\tb = 1.6$ or $40$.
}
\label{fig:mh_mq}
\end{figure}

\smallskip
\reffi{fig:mh_mq} shows the dependence of $\mh$ on $\msq$ for the cases
of no mixing and maximal mixing, and we have set $\MA = 500 \gev$.
Very good agreement is found  in the
no-mixing scenario as well as in the maximal-mixing scenario,
the deviation lies below $2 \gev$.

\begin{figure}[ht!]
\begin{center}
\mbox{
\psfig{figure=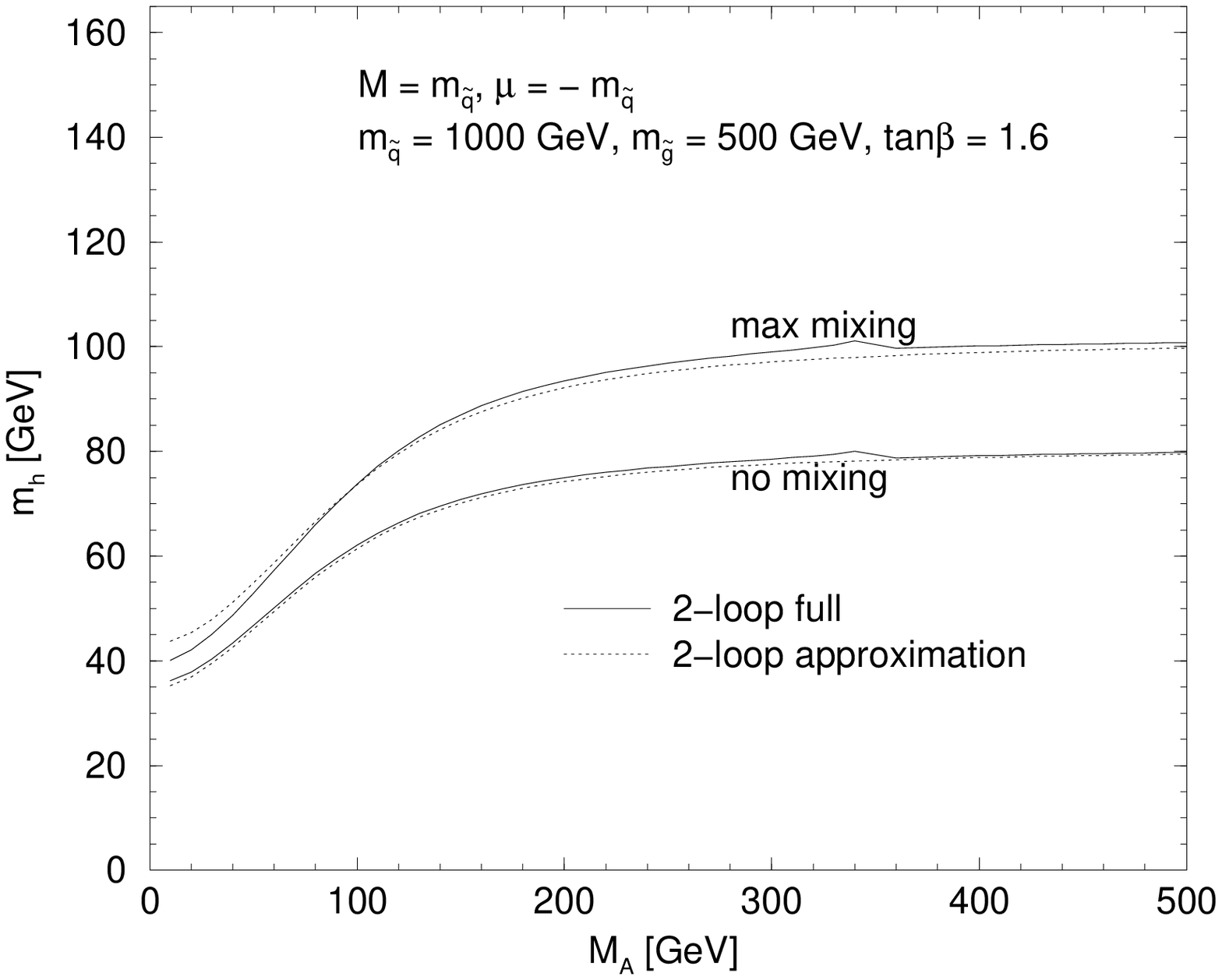,width=7cm,height=7cm}}
\hspace{1em}
\mbox{
\psfig{figure=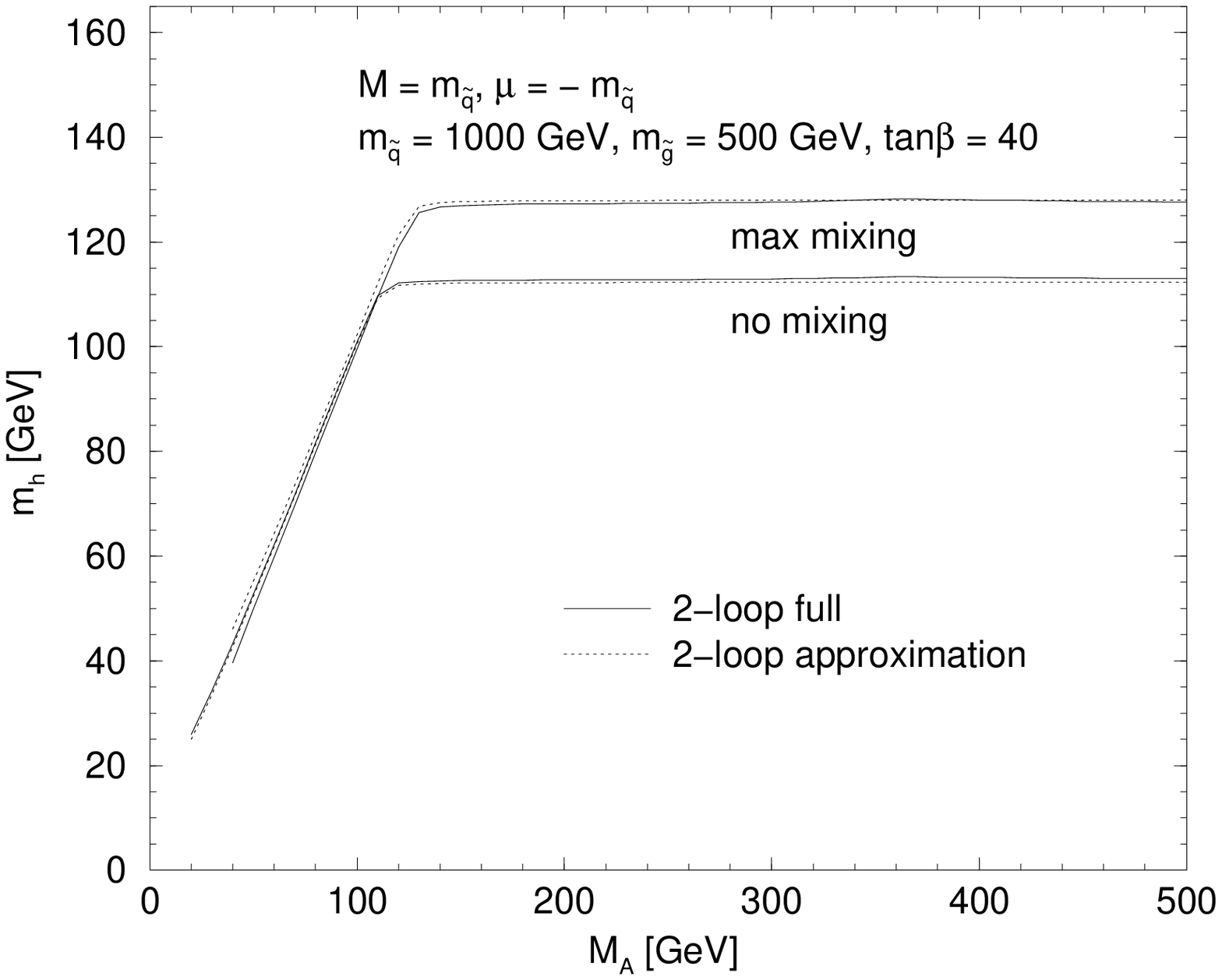,width=7cm,height=7cm}}
\end{center}
\vspace{-1.5em}
\caption[]{
$\mh$ as a function of $\MA$, calculated from the full formula and
from the approximation formula for $\msq = 1000 \gev, 
\mgl = 500 \gev$ and $\tb = 1.6$ or $40$.
}
\label{fig:mh_MA}
\end{figure}


\smallskip
The dependence on $\MA$ is shown in \reffi{fig:mh_MA}. 
The quality of the approximation is typically better than $1 \gev$ for
the no-mixing case and better than $2 \gev$ for the maximal-mixing
case. Only for very small (and experimentally already excluded) values
of $\MA$ a deviation of $5 \gev$ occurs. 
The peaks in the plot for $\tb = 1.6$ in the full result are due to
the threshold  $\MA = 2\,\mt$ in the
\onel\ contribution, originating from the top-loop diagram in
the $A$~self-energy. This peak does not occur in the approximation
formula and can thus lead to a larger deviation around the
threshold.

\smallskip
In deriving the leading and subleading corrections we have set 
$\mgl = \sqmsmt$, thus eliminating this additional scale. A~variation of
$\mgl$ in the full result directly corresponds to a shift relative to the
approximation formula. As it was shown in \citere{mhiggslong}, this
deviation is negligible for the no-mixing case and lies within 
$\pm 2 \gev$ in the maximal-mixing case.
It was also analyzed in \citere{mhiggslong} that the variation of $\mh$
with  
the $SU(2)$ gaugino mass parameter $M(=M_2)$ and with  the
Higgs-mixing parameter $\mu$ is relatively weak. We have found only small
deviations of 
the approximation formula with respect to the full result,
not more than $2 \gev$ for small values of $M$, 
and $1.5 \gev$ for small values of $\mu$.

\begin{figure}[ht!]
\begin{center}
\hspace{1em}
\mbox{
\psfig{figure=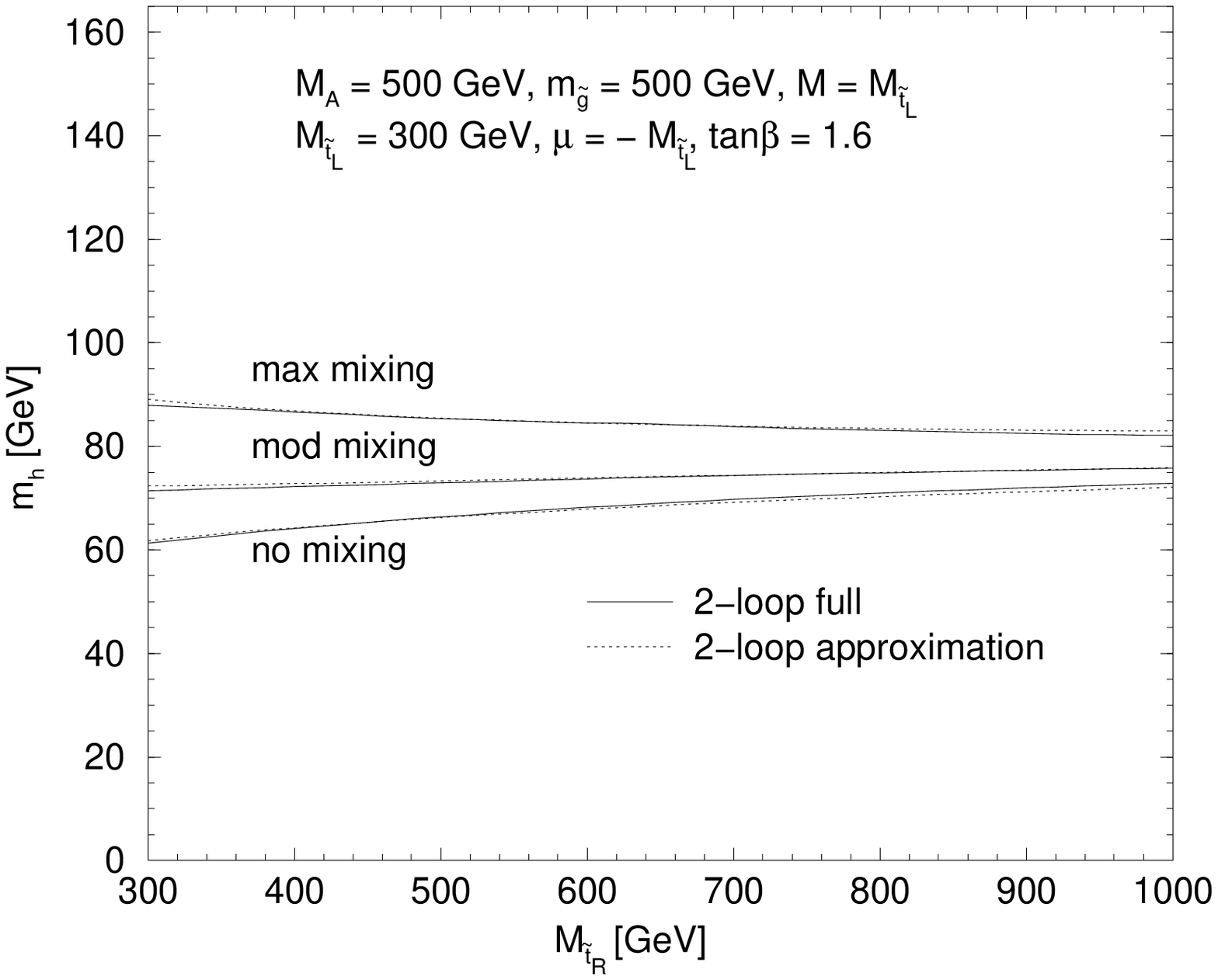,width=7cm,height=7cm}}
\hspace{1em}
\mbox{
\psfig{figure=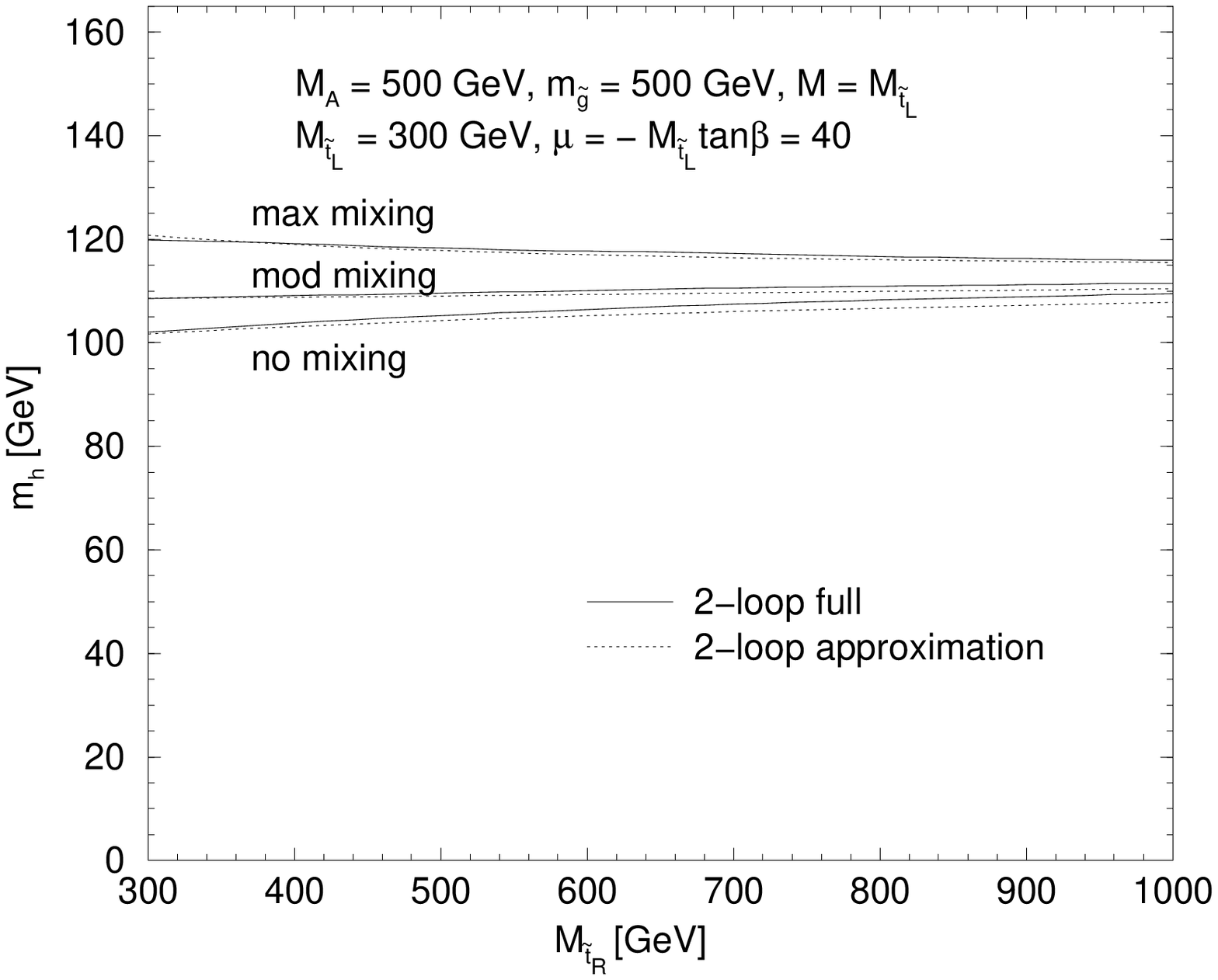,width=7cm,height=7cm}}
\end{center}


\begin{center}
\hspace{1em}
\mbox{
\psfig{figure=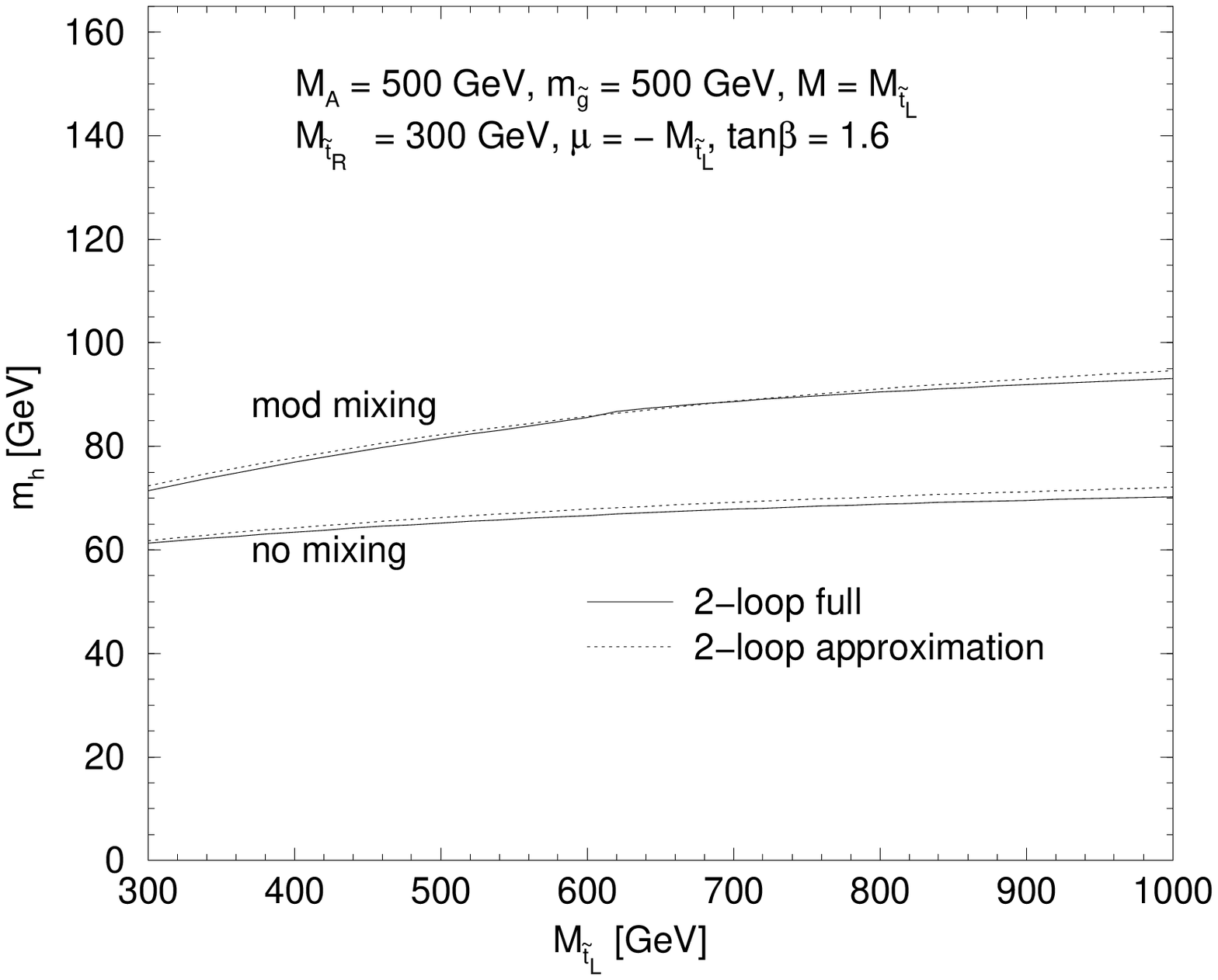,width=7cm,height=7cm}}
\hspace{1em}
\mbox{
\psfig{figure=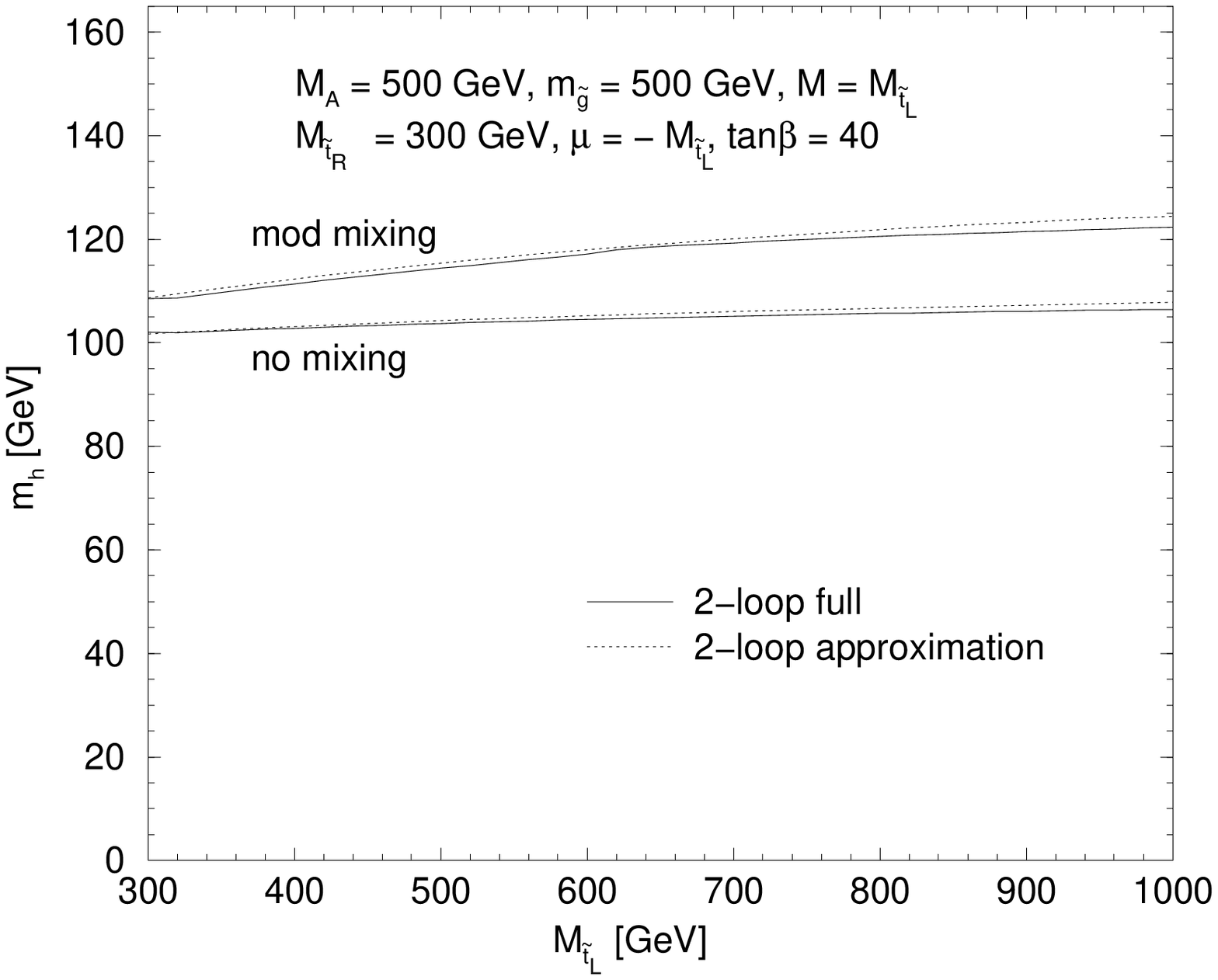,width=7cm,height=7cm}}
\end{center}
\vspace{-1.5em}
\caption[]{
$\mh$ as a function of $\MstR$ or $\MstL$, 
calculated from the full formula and
from the approximation formula for varied $\MstR~(\MstL)$,
$\MstL~(\MstR) = 300 \gev$,
$\MA = 500 \gev, \mgl = 500 \gev$ and $\tb = 1.6$ or $40$.
} 
\label{fig:mh_MstLR}
\vspace{1cm}
\end{figure}

\smallskip
Finally we consider the case when $\MstL \neq \MstR$, which is shown
in \reffi{fig:mh_MstLR}. Here also the case $\Mtlr = \MstL$ 
is depicted, which we refer to as 'moderate mixing'. 
'No mixing' here corresponds to $\Mtlr = 0 $, 'maximal mixing'
corresponds to the 
choice $\Mtlr = 2\;\MstL$. For fixed $\MstR$ this leads to 
a $\Stop$-mass below the experimental lower bound
for nearly the whole parameter space; thus we have omitted
this case in this scenario (bottom of \reffi{fig:mh_MstLR}). Here we
have made use of \refeq{msfkt}. Agreement better than $2 \gev$ is found
for all scenarios.


\section{Conclusions}

By means of a Taylor expansion of the diagrammatic \twol\ result
we have derived the leading logarithmic and non-logarithmic as well as 
subleading contributions to the lightest Higgs-boson mass
$\mh$ up to $\oaas$. 
This result has been incorporated into a compact analytical
formula for $\mh^2$, which can easily be implemented for numerical
evaluation. It contains the tree-level expression, leading and
subleading \onel\ contributions, the leading logarithmic and
non-logarithmic \twol\ contributions, and two further corrections beyond
$\oaas$.
This formula is valid for general mixing in the scalar top sector and 
arbitrary choices of the parameters in the Higgs sector of the model.
It has been included into the Fortran codes \fh\ and \fhf,
thus allowing a very fast and reasonably accurate evaluation of $\mh$. 
The codes are available via the WWW page\\
{\tt http://www-itp.physik.uni-karlsruhe.de/feynhiggs}~.

The approximation formula has been compared with the full
diagrammatic result, in which the complete \onel\ and the 
dominant \twol\ corrections of $\oaas$ are taken into account without 
Taylor expansion. 
Good agreement, better than $2 \gev$, is found for most parts 
of the MSSM parameter space. Larger deviations, exceeding a few GeV,
can occur for large mixing in the $\Stop$-sector,
$|\Mtlr/\msq| \gsim 2$. 
The effect of the corrections of $\oaas$ of shifting the maxima for
$\mh$ towards smaller values of $|\Mtlr/\msq|$ can easily be read off
from the compact formula by analytically determining the extrema of $\mh$.



\bigskip
\subsection*{Acknowledgements}
We thank H.~Haber and C.~Wagner for helpful discussions.
W.H. gratefully acknowledges support by the Volkswagenstiftung.




\end{document}